\documentclass[review]{elsarticle}

\usepackage{lineno}
\usepackage{hyperref}

\usepackage{makeidx}  
\usepackage{url}
\usepackage{graphicx}
\usepackage{amssymb}
\usepackage{pifont}
\usepackage{latexsym}

\newcommand{\nop}[1]{}

\newcommand{\term}[1]{{\tt #1}}

\newtheorem{example}{Example}

\input{psfig}

\modulolinenumbers[5]

\begin{document}

\begin{frontmatter}

\title{Stream Reasoning on Expressive Logics}


\author[mymainaddress]{Gulay Unel}
\ead{gulay.unel@isikun.edu.tr}


\address[mymainaddress]{Department of Information Technologies, I\c{s}{\i}k University, \c{S}ile, {\.I}stanbul, Turkey 34980}

\begin{abstract}
Data streams occur widely in various real world applications.
The research on streaming data mainly focuses on the
data management, query evaluation and optimization on these data, however
the work on reasoning procedures for streaming knowledge bases on both the
assertional and terminological levels is very limited.
Typically reasoning services on large knowledge bases are very expensive, and need to
be applied continuously when the data is received as a stream.
Hence new techniques for optimizing this continuous process is needed for developing
efficient reasoners on streaming data.
In this paper, we survey the related research on reasoning on expressive logics that can be applied to
this setting, and point to further research directions in this area.
\end{abstract}


\end{frontmatter}


%
%
%
%
%
 you are submitting to.
%
%
%
\input{psfig}

\section{Introduction}
\label{sec:introduction}
Reasoning is typically applied to static knowledge bases and often ignored for rapidly changing data at both the
terminological and assertional levels. However, there is a clear need for the design and implementation of
reasoning methods for dynamic knowledge bases motivated by the increase in the use of sensor data, streaming web and multimedia data,
rapidly changing medical and financial data etc.
In addition to the need on reasoning over streaming data, there are also
applications where complex reasoning tasks need to be continuously applied such as verifying ontologies using ontology editors,
service discovery and matchmaking on web service frameworks
where services register and deregister rapidly, and logical learning from fluctuating data.

The research on streaming data mainly focuses on the
data management and query processing~\cite{Chaudhry,Golab03}, however the work on reasoning procedures which involves deducing new information
from what is known is very limited.
Data streams are widely available, however their usage is mainly restricted to retrieval and search, instead of
tasks such as decision making and
deriving conclusions from them which can be used in many applications.
The information searched and uploaded to Web which has a very dynamic nature can be formalized with the advances on Semantic Web
which in turn results in the ability of applying reasoning methods over this formalization.
For instance, the spread of a health disease can be detected by reasoning over the streaming
keywords and/or information uploaded to the Web.

The reasoning techniques on streaming knowledge bases need to support incrementality to be able to use the results of the previous
computations in the active reasoning step performed after the arrival of new data. In addition, soundness and
completeness is not crucial for some applications. Hence, more efficient
incremental reasoning techniques without the requirement of soundness/completeness can be used in such applications.

In this paper, we provide relevant research on incremental reasoning techniques
on expressive logics: RDFS and OWL, Description Logics, rule-based Logics, Description Logic Programs (DLP), other fragments of first order and second order logics. We will also discuss hybrid approaches.

%


The paper is structured as follows:
Section~\ref{sec:problem} defines the problem of stream reasoning.
Section~\ref{sec:lres} describes the logics and reasoning problems.
Finally, the conclusions and future research directions are given in Section~\ref{sec:conc}.

\section{Stream Reasoning: Problem Statement}
\label{sec:problem}
We define stream reasoning as the problem of generating a stream of conclusions from reasoning
over terminological and/or assertional axioms.
Main components of this problem are:
\begin{itemize}
\item A stream of terminologies: $T_1,$ $T_2,$ $\ldots,$ $T_n$
\item A stream of assertions: $A_1,$ $A_2,$ $\ldots,$ $A_m$
\item A stream of conclusions: $C_1,$ $C_2,$ $\ldots,$ $C_k$
\end{itemize}
The reasoning problem is defined as given an ordered set $\{T_1,$
$T_2,$ $\ldots$ $T_n\}$ and/or $\{A_1,$ $A_2,$ $\ldots$ $A_m\}$
determine whether they entail certain ordered set of conclusions
$\{C_1,$ $C_2,$ $\ldots$ $C_k\}$. $C_i$ is entailed by $\{T_1,$
$T_2,$ $\ldots$ $T_n\}$ and/or $\{A_1,$ $A_2,$ $\ldots$ $A_m\}$ if
it is entailed by the terminologies and/or assertions that come
before $C_i$ in terms of the time parameter provided by the
streaming setting.

The naive approach to solve this problem is to apply the reasoning
procedure from scratch whenever the knowledge base is changed which
is typically very expensive. Hence there are approaches proposed to
perform this process incrementally. These approaches are also based
on different `continuous' semantics applicable to the particular
logical formalism used.




\section{Logics and Reasoning Problems}
\label{sec:lres}
\begin{figure}[t]
\centering
\includegraphics [width=100mm]{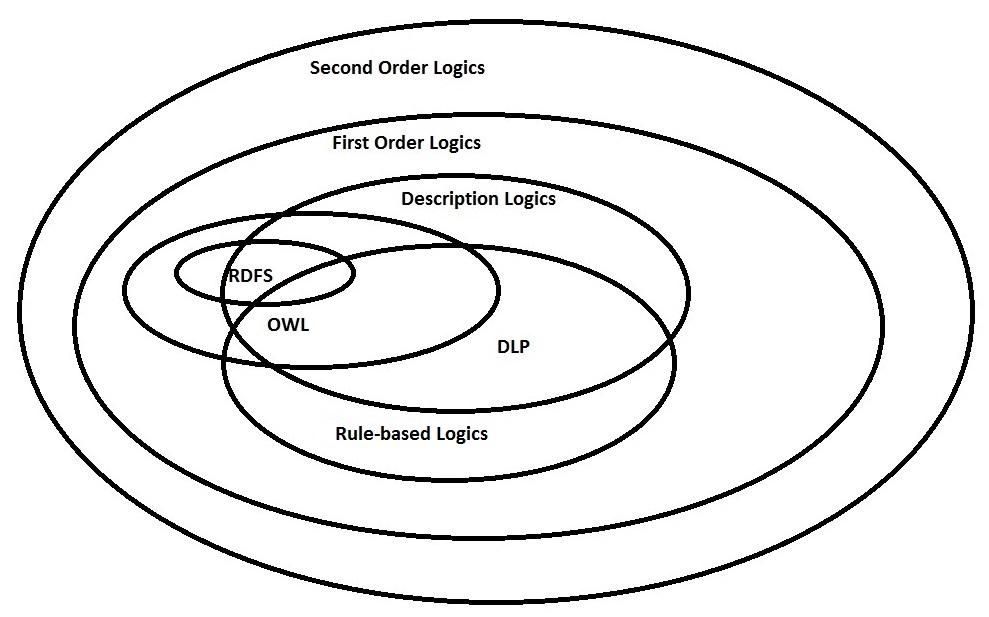}
\caption{Taxonomy of the logics}
\label{fig:taxonomy}
\end{figure}

In this section we provide an outlook on the current state of expressive stream reasoning including the logics and reasoning problems. The logics
we investigate are RDFS and OWL, Description Logics, rule-based Logics, Description Logic Programs, and other fragments of First Order and Second Order Logics as shown in Figure~\ref{fig:taxonomy}. We also provide results
on the hybrid approaches.

\subsection{RDFS and OWL}
\label{ss:rdfsowl}

Most of the work on stream reasoning on Semantic Web formalisms focus on RDF streams. In this section, we provide stream reasoning approaches on RDFS and OWL streams.

In \cite{Urbani13} incremental materialization of $\rho$df RDFS fragment is presented which is implemented as a parallel system. The performance is evaluated on a
prototype system called {\it DynamiTE}. The evaluation results show that the system is able to compute the materialization of knowledge bases up to one million statements
and can perform updates of two hundred thousand triples in a range from hundred of milliseconds to less than two minutes.

In \cite{Ren11} an ontology stream reasoning approach for OWL 2 EL based on Truth Maintenance Systems (TMS) is presented which is called an ontology stream management system (OSMS).
The evaluation shows that the approach performs better than naive re-computation with a knowledge base up to 30000 axioms and updates up to 10\%. The work on OWL DL is presented in Section \ref{ss:dl}.

\subsection{Description Logics}
\label{ss:dl}
The stream reasoning research in Description Logics includes methods proposed
for incremental A-Box \cite{Wiener06} and T-Box \cite{Parsia06,Grau07}
reasoning. The following reasoning problem
based on an example given in \cite{Grau07} demonstrates
relevant aspects of incremental reasoning.

\newcommand{\indentruledl}{\hbox to 5mm{\hfil}}

\begin{example}\label{ex:dl}
Consider the following ontology $O^1$:

\vskip2mm
\noindent
$\indentruledl D1:$ $Cystic\_Fibrosis \equiv Fibrosis \sqcap \exists located\_In.Pancreas$\\
$\indentruledl D2:$ $Pancreatic\_Fibrosis \equiv Fibrosis \sqcap Pancreatic\_Disorder$\\
$\indentruledl S1:$ $Pancreatic\_Disorder \sqsubseteq Disorder \sqcap \exists located\_In.Pancreas$\\

\noindent
The following conclusions can be drawn from  $O^1$:
\vskip2mm
\noindent
$\indentruledl C1:$ $Pancreatic\_Fibrosis \sqsubseteq Cystic\_Fibrosis$\\
$\indentruledl C2:$ $Pancreatic\_Fibrosis \sqsubseteq Disorder$\\
\vskip2mm

\noindent
$C1$ follows from $D1,$ $D2$, and $S1$. $C2$ follows from $D2$ and $S1$.
Now consider an update on the axiom $D1$ of $O^1$ which results in a modified ontology $O^2$:

\vskip2mm
\noindent
$\indentruledl D1^\prime:$ $Cystic\_Fibrosis \equiv Fibrosis \sqcap \exists located\_In.Pancreas \sqcap \\
{\hbox to 40mm{\hfil}} \exists has\_Origin.Genetic\_Origin$\\

\noindent
The conclusion $C1$ now becomes:

\vskip2mm
\noindent
$\indentruledl \neg (C1^\prime:$ $Pancreatic\_Fibrosis \sqsubseteq Cystic\_Fibrosis)$\\
\vskip2mm

\noindent
Since we know that $C1$ follows from $D1$, $D2,$ and $S1$ when any of these change we need to check if
$C1$ still hold, however since $C2$ does not depend on $D1$ we can conclude that it will still hold after the update.
\end{example}



The work on incremental reasoning on dynamic A-Boxes~\cite{Wiener06} investigates the process of
updating tableau completion graphs used for consistency checking incrementally.
This technique is proposed for $\mathcal{SHOQ (D)}$ and $\mathcal{SHIQ (D)}$ which
correspond to a large subset of OWL-DL. The presented algorithm for updating completion graphs
considers both addition and removal of A-Box assertions where the addition case is
straightforward due to the incremental and non-deterministic nature of the tableau algorithm.
An empirical analysis of the algorithm through an implementation in the Pellet\footnote{http://clarkparsia.com/pellet/}
reasoner is also provided.
The presented results demonstrate orders of magnitude performance improvement.

Related T-Box reasoning techniques include incrementally updating classifications of
ontologies encoded in $\mathcal{SHOIN (D)}$ which also corresponds to OWL-DL \cite{Parsia06}.
Classification is the determination for every two named classes $A$ and $B$ in an ontology
whether $A$ subsumes $B$ and vice versa. Classification involves $n^2$ subsumption tests for
$n$ named classes in the worst case. Since subsumption tests are typically very expensive, optimization
techniques for classification involves avoiding unnecessary tests and substituting
cheap tests (sound but possibly incomplete) whenever possible. The techniques proposed
for incremental reasoning includes the utilization of these optimizations to the streaming setting
where there are axiom additions and deletions.
The optimization on avoiding tests involve maintaining a subsumption hierarchy and whenever a new axiom is added
using this information to reduce the number of subsumption tests needed for the classification after the update.
This technique is independent from the reasoning algorithm used, hence it can be used for incremeantal reasoning
in other logical formalisms.
The second optimization on using cheaper tests uses the notion of model caching~\cite{Horrocks97}
(or pseudo model caching as in~\cite{Haarslev01}). The general idea is to build pseudo models
during the satisfiability checking of classes which is done prior to classification.
These models can be reused in subsumption checking. If a model of a class $A$ and a model of a class
$\neg B$ can be combined into a a single model then their conjunction is satisfiable which also
means $A$ does not subsume $B$. Hence whenever a new axiom is added to the T-Box the previously built
pseudo models can be used for incremental classification. This approach is sound but incomplete.
Since most subsumption tests are negative, it can dramatically improve the classification performance.
The empirical analysis of the mentioned techniques using the Pellet
reasoner  \cite{Parsia06} also show promising results.

The second incremental reasoning technique for $\mathcal{SHOIN (D)}$ T-Boxes~\cite{Grau07} also involves reusing of information
obtained from previous versions of a T-Box. The proposed method is based on the notion of a module and can be
applied to arbitrary queries against ontologies expressed in OWL-DL.
The general idea is to keep track of evidences for both subsumption and non-subsumption relationships
to be reused after an update to the ontology. For the classification of subsumptions after an update,
the axioms that each subsumption follows from is stored and if the update does not change
any of these axioms it is guaranteed that the subsumption will still hold after the update.
For determining the entailment of non-subsumtions, again an evidence can be used such as a counter model
that is constructed by tableaux-based procedures which can be reused for reasoning about non-subsumtions
after a T-Box update as also described in \cite{Parsia06}. The evaluation of the proposed methods
demonstrate substantial speed-up when compared to regular classification.

More recently, incremental reasoning approaches based on module extraction on $\mathcal{SROIQ}$ have been proposed \cite{Sirin07,Grau2010}.
Evaluation of these algorithms to solve the classification logical inference problem shows reasonable response times \cite{Reyes2014}.

\subsection{Rule-based Logics}
\label{ss:horn}
In this section, we will outline relevant work on rule-based logics.
Early work is mainly based on Default Logic~\cite{Reiter80}.
The main reason for this is that Default Logic can capture the dynamic nature of the reasoning process and derived
conclusions better than other rule based logical formalisms which we will describe in further detail later in this section.
The following example shows how a rule-based evaluation problem can be applied on a streaming setting.

\newcommand{\indentrule}{\hbox to 25mm{\hfil}}

\begin{example}\label{ex:reach}
Consider we have connectivity information on a set of Web pages which is updated frequently
as new pages enter or leave and a program computing reachability among these Web pages
represented by the following program:

\vskip2mm
\noindent
$\indentrule connected(w_1,w_2).$ \\
$\indentrule connected(w_2,w_3).$ \\
$\indentrule \ldots$ \\
$\indentrule \ldots$ \\
$\indentrule \ldots$ \\
\vskip2mm
\noindent
$\indentrule reachable(W,Z) \leftarrow connected(W,Z).$ \\
$\indentrule reachable(W,Z) \leftarrow reachable(W,Y), connected(Y,Z).$ \\
\vskip2mm

\noindent
We need efficient procedures for maintaining the reachability information and answering queries on the program as
new data arrives and leaves. For instance we should not recompute the fact $reachable(w_1,w_2)$ in the reasoning process
after the arrival of the fact $connected(w_2,w_3)$.
\end{example}



In addition to the streaming nature of knowledge, the reasoning itself can also be considered as an
ongoing process deducing a stream of conclusions over time rather than focusing on the end results.
This adds the notion of time to the reasoning process. In~\cite{Elgot88},
Step Logic is introduced as a formalism which is used to characterize this ongoing process of
reasoning. In this setting, the reasoner starts out with an empty set of beliefs at time zero and
certain conclusions or observations may arise at discrete time steps. Hence, they can
represent streaming knowledge and conclusions. A subset of the overall knowledge base is the
focus of attention at a certain time which also adds a new dimension, namely space, to the reasoning process.
Conclusions can be drawn from the current set of beliefs and also can be retracted when additional knowledge
arrives. The inconsistencies can be dealt by tagging rules and facts based on their source such as
observation or deduction, and also by considering the time parameters of the contradictory facts
that are generated. A description of such a system is given in~\cite{Elgot91}. Efficient
reasoning techniques for such a system that reuses its previous deduction steps and
selects the relevant parts of the knowledge base for reasoning still needs to be explored.

Active Logic~\cite{Elgot99} introduced after Step logic also considers reasoning as an ongoing process
altering plans and beliefs accordingly. Similarly a uniform, lightweight language specification for
a class of temporal logics for reasoning about action and change is described in~\cite{Doherty98}.

The following example demonstrates the use of a non-monotonic reasoning problem on a streaming knowledge base.

\begin{example}\label{ex:default}
Consider the following program (terminology),

\vskip2mm
\noindent
$\indentrule flies(X) \leftarrow bird(X).$ \\
$\indentrule \neg flies(X) \leftarrow ostrich(X).$ \\
\vskip2mm

\noindent
and the streaming fact base (assertions):

\vskip2mm
\noindent
$\indentrule bird(tweety).$ \\
$\indentrule ostrich(tweety).$ \\
\vskip2mm

\noindent
Note that the fact $flies(tweety)$ is derived from the fist rule, and as new information revealing
the fact $ostrich(tweety)$ arrives $\neg flies(tweety)$ is derived which contradicts with the
previous conclusion. Since $\neg flies(tweety)$
is more recent (deduced after $flies(tweety)$ when the new information $ostrich(tweety)$ arrived to the fact base)
the previous one can be ignored which
can be implemented naturally in the streaming setting since it has a time dimension.
\end{example}



Work on constraint
programming architectures \cite{Robin07} outlines the desirable
properties of such platforms which also include the solution
adaptation. The availability of this property among others is in
several constraint programming platforms is noted and a new
architecture with this property is proposed. The main solution
adaptation technique mentioned is {\it Justification Based Truth
Maintenance} where each solution is kept with the indexes of the
rules that it is derived from which is similar to the incremental
techniques we outlined for Description Logics. The work on solving
database integrity constraints \cite{Henning06} aims for producing
simplified incremental checks for each database update.

There is also work on programming paradigms such as {\it Finite Differencing} \cite{Paige90}
which is an old mathematical idea that can speed up computation by replacing repeated costly calculations
by more efficient counterparts. In addition to this, Dynamic Logic \cite{Harel84} is proposed as a formal
system for reasoning about programs. The name {\it Dynamic Logic} emphasizes the principal feature
distinguishing it from classical predicate logic. In the latter, the truth is static.
In Dynamic Logic, there are explicit syntactic constructs called programs whose main role is to change
the values of variables, thereby changing the truth values of programs.

Furthermore, there is work on belief revision \cite{Darwiche96,Eijck08} which is the process of changing beliefs to
take into account new pieces of information, where one method is the use of Default Logic explained in this paper.
Work on interpreting and summarizing (i.e. compressing) streaming factual data
is also relevant to stream reasoning which involves techniques form
logical formalisms such as Inductive Logic Programming \cite{Muggleton94}.

After the introduction of stream reasoning \cite{SR2009intro} new techniques on various formalisms emerged. One of them
is Stream reasoning on Answer Set programming (ASP) \cite{Gebser12} which provides
language constructs and modeling techniques for reasoning on time-decaying logic programs. The stream reasoning
capability is incorporated to the logic programming language with additional ASP constructs for the emergence and expiration of program parts.

ETALIS - a system that performs stream reasoning on top of Complex Event Processing (CEP) is implemented later
which provide real-time specification and monitoring of changes \cite{ARFS2012}. The system improves CEP by performing reasoning on streams combined with background knowledge on-the-fly using a deductive rule-based paradigm.
Event-driven processing is included which enables the system to derive a complex event at the moment it occurs.
An extension of ETALIS that accepts RDFS ontologies as background knowledge which are converted to Prolog programs
is also described. The system is implemented in Prolog and has been used in research projects.

Stream reasoning on a temporal rule-based formalism called LARS \cite{BeckDE15} with semantics similar to Answer Set Programming is also a recent work in the area. The work presented provides an incremental answer update algorithm on a fragment of LARS programs based on stream stratification which extends techniques from truth maintenance systems.

An incremental update algorithm for Datalog materialisation is proposed in \cite{Motik15}. The algorithm is shown to extend the
materialisation techniques especially for data that is subject to large and/or frequent changes.

\subsection{DLP}
\label{ss:dlp}

There is a system developed in Haskell using plausible Description Logic Programs
where the main components are description logic, plausible logic and functional programming \cite{Groza12}.
The proposed system aims handling theoretically infinite and heterogenous data streams by lazy evaluation
which is characterized as a step towards building real-time stream reasoners.

\subsection{Other Fragments of First Order and Second Order Logics}
\label{ss:higher}

Recently, a technique was proposed for checking the satisfiability of Monadic second-order formulas incrementally
which can be used for reasoning on a series of formulas on large theories, such as
those corresponding to rapidly changing constraints encoded with expressive logics \cite{Unel2015}. The method
is implemented using logic programming which is an application of incremental reasoning over
a more expressive logic.

\subsection{Hybrid Approaches}
\label{ss:hybrid}
In recent years research towards stream reasoning systems provided architectures and results.
Firstly, a stream reasoning architecture for handling different complexity levels by the notion of cascading reasoners
was proposed \cite{Heiner10}. The architecture is designed to provide a wide array of reasoning capabilities from
raw stream processing to complex Description Logic reasoning.

\subsection{Summary}
The different logics and reasoning methods we focused on related to stream reasoning are summarized in Table \ref{tab:comp} which also includes whether the work is implemented or not.
The reasoning problems we consider in our examination are
classification, materialization (materialized derivations), satisfiability, query answering, entailment and consistency.
Different dialectics of Description Logics are shown in the summary. Logics that are composed of rules are generalized as rule-based logics.
The summary of the results show that the work on Description Logics mostly examine the problem of consistency whereas the work on rule-based logics focus on the problem of query answering.
The work on RDFS and OWL, DLP and other fragments of First Order and Second Order Logics is very limited compared to rule-based and Description Logics.

Most recent stream reasoning approaches based on Description Logics show that the incremental classification problem can be solved efficiently \cite{Sirin07,Grau2010,Reyes2014}.
One of the most recent promising works on rule-based logics is on Answer Set programming (ASP) \cite{Gebser12} where stream reasoning
is incorporated to the logic programming language. The preliminary experiments of this work show that there is still need for re-structuring the rule-based systems 
according to stream reasoning requirements.

\begin{table}[t]
\centering
\caption{Summary of the results}
\begin{tabular}{ | l |l | l | l |}
\hline
Logic  & Reasoning Problem(s)& Impl. & Reference(s) \\
\hline
RDFS fragment  $\rho$df &Materialization &$\checkmark$ &\cite{Urbani13}  \\
 & Query Answering&  &   \\
OWL 2 EL &Entailment &$\checkmark$ &\cite{Ren11}  \\
$\mathcal{SHOQ (D)}$/$\mathcal{SHIQ (D)}$ & Consistency&$\checkmark$ &\cite{Wiener06}  \\
$\mathcal{SHOIN (D)}$ & Classification& $\checkmark$ & \cite{Parsia06} \\
$\mathcal{SHOIQ}$ & Classification& $\checkmark$  & \cite{Grau07}  \\
$\mathcal{SROIQ}$ & Classification& $\checkmark$  & \cite{Sirin07,Grau2010,Reyes2014}  \\
Step/Active Logic& Query Answering&$\checkmark$ &\cite{Elgot88,Elgot99} \\
Rule-based Logics& Consistency& \ding{55}& \cite{Henning06} \\
Rule-based Logics& Entailment& $\checkmark$ & \cite{Gebser12} \\
 & Query Answering&  &   \\
Rule-based Logics& Query Answering& $\checkmark$ & \cite{ARFS2012} \\
Rule-based Logics& Query Answering& \ding{55} & \cite{BeckDE15} \\
Rule-based Logics& Materialization& $\checkmark$ & \cite{Motik15} \\
DLP& Query Answering& $\checkmark$ & \cite{Groza12} \\
Monadic Second Order& Satisfiability& $\checkmark$ & \cite{Unel2015} \\
Logics & &  &   \\
\hline
\end{tabular}
\label{tab:comp}
\end{table}

\section{Conclusions and Future Work}
\label{sec:conc}
There is a gap between the research on advanced reasoning techniques and reasoning on streaming data.
In this paper, we provided related work on the existing reasoning methods that can be applied
to streaming knowledge bases. This presented results illustrate that we can develop optimized decision procedures on streaming knowledge bases even if they are expressed in logics with complex decision problems.

%
%

Considering the amount of streaming information in current Web
which will continue increasing, there is a clear need for developing
logical frameworks for stream reasoning which incorporates different reasoning
techniques for various logical formalisms ranging from light weight to more expressive ones.
Such platforms can make use of various evaluation and optimization methods
for stream reasoning and can provide customized reasoning procedures for different application settings.
Various related parameters to be considered for such systems are
{\it heterogeneity, time and window dependencies, noisy and uncertain data, scale, real-time constraints,
continuous processing, distribution of computational units} as outlined in \cite{SR2009intro}.

Furthermore, considering the results of the analysis of current relevant techniques, it can be concluded that
efficient algorithms applicable to stream reasoning on various formalisms are currently
under development. After the introduction of stream reasoning as a target research area 
this development is accelerated and currently there are systems that are being used in various applications.



\bibliography{srlogic}

\end{document}